\documentclass[twocolumn]{aastex63}
\usepackage{savesym}
\usepackage{graphicx}	
\usepackage{amsmath}	
\usepackage{amssymb}	
\usepackage[super]{nth}
\savesymbol{tablenum}
\usepackage{siunitx}
\usepackage[version=4]{mhchem} 
\usepackage{lineno}
\newcommand{\solar}[1]{\text{#1}_\odot}

\DeclareSIUnit\parsec{pc}
\DeclareSIUnit\lightyear{ly}
\DeclareSIUnit\erg{erg}
\DeclareSIUnit\yr{yr}
\DeclareSIUnit\au{AU}
\DeclareSIUnit\jansky{Jy}
\DeclareSIUnit\sunM{\solar{M}}
\DeclareSIUnit\sunL{\solar{L}}
\DeclareSIUnit{\solarlum}{L_\Sun}

\submitjournal{ApJ}

\shorttitle{The Growth of the ICL in SpARCS1049+56}
\shortauthors{Barfety et al.}
\graphicspath{{./}{figures/}}

\begin{document}

\title{An Assessment of the In-Situ Growth of the Intracluster Light  in the High Redshift Galaxy Cluster SpARCS1049+56 }

\correspondingauthor{Capucine Barfety}
\email{capucine.barfety@mail.mcgill.ca}

\author[0000-0002-1952-3966]{Capucine Barfety}
\author{F\'elix-Antoine Valin}
\author[0000-0002-0104-9653]{Tracy M.A. Webb}
\affiliation{Department of Physics, McGill Space Institute, McGill University, 3600 rue University, Montr\'eal, Qu\'ebec, Canada, H3A 2T8}

\author[0000-0001-7095-7543]{Min Yun}
\affiliation{Department of Astronomy, University of Massachusetts Amherst, Lederle Graduate Research Tower, 619-E, 710 N Pleasant Street, Amherst, MA 01003-9305}

\author{Heath Shipley}
\affiliation{Department of Physics, McGill Space Institute, McGill University, 3600 rue University, Montr\'eal, Qu\'ebec, Canada, H3A 2T8}

\author[0000-0002-5828-6211]{Kyle Boone}
\affiliation{DiRAC Institute, Department of Astronomy, University of Washington, 3910 15th Ave NE, Seattle, WA, 98195, USA}

\author{Brian Hayden}
\affiliation{Space Telescope Science Institute, 3700 San Martin Drive Baltimore, MD 21218, USA}

\author[0000-0001-7271-7340]{Julie Hlavacek-Larrondo}
\affiliation{D\'epartement de physique, Universit\'e de Montr\'eal, Complexe des Sciences, Case postale 6128, succursale Centre-ville, Montr\'eal, Qu\'ebec, H3C 3J7, Canada}

\author[0000-0002-9330-9108]{Adam Muzzin}
\affiliation{Department of Physics and Astronomy, York University, 4700 Keele St., Toronto, Ontario, Canada, MJ3 1P3}

\author[0000-0003-1832-4137]{Allison G. Noble}
\affiliation{ASU School of Earth and Space Exploration, PO Box 871404, Tempe, AZ 85287, USA}

\author[0000-0002-4436-4661]{Saul Perlmutter}
\affiliation{Department of Physics, University of California, 366 Physics North, MC 7300, Berkeley, CA, 94720-7300}

\author[0000-0003-2001-1076]{Carter Rhea}
\affiliation{D\'epartement de physique, Universit\'e de Montr\'eal, Complexe des Sciences, Case postale 6128, succursale Centre-ville, Montr\'eal, Qu\'ebec, H3C 3J7, Canada}

\author{Gillian Wilson}
\affiliation{Department of Physics and Astronomy, University of California-Riverside, 900 University Avenue, Riverside, CA 92521, USA}

\author[0000-0003-4935-2720]{H.K.C. Yee}
\affiliation{The David A. Dunlap Department of Astronomy and Astrophysics, University of Toronto, 50 St George St., Toronto, ON M5S 3H4,
Canada}



\begin{abstract}

 The formation of the stellar mass within galaxy cluster cores is a poorly understood process.  It features the complicated physics of cooling flows, AGN feedback, star formation and more. Here, we study the growth of the stellar mass in the vicinity  of the Brightest Cluster Galaxy (BCG) in a $z=1.7$ cluster, SpARCS1049+56. We synthesize a reanalysis of existing HST imaging, a previously published measurement of the star formation rate, and the  results of new radio molecular gas spectroscopy. These analyses represent the past, present and future  star formation respectively within this system. We show that a large amount of stellar mass -- between $(2.2 \pm 0.5) \times 10^{10} \: M_\odot$ and $(6.6 \pm 1.2) \times 10^{10} \: M_\odot$ depending on the data processing -- exists in a long and clumpy tail-like structure that lies roughly 12 kpc off the BCG. Spatially coincident with this stellar mass is a similarly massive reservoir (1.0$\pm$0.7$\times$10$^{11}$M$_\odot$)  of molecular gas that we suggest is the fuel for the immense star formation rate of 860 $\pm$ 130 M$_\odot$yr$^{-1}$, as  measured by infrared observations. \citet{hl2020} surmised that massive, runaway cooling of the hot intracluster X-ray gas was feeding this star formation, a process that had not been observed before at high-redshift.  We conclude, based on the amount of fuel and current stars, that this event may be rare in the lifetime of a cluster, producing roughly 15 to 21\% of the Intracluster Light (ICL) mass in one go, though perhaps a common event for all galaxy clusters. 
 
 
\end{abstract}



\section{Introduction} \label{sec:intro}
The centres of galaxy clusters remain among the most violent yet least understood regions of the universe, at all redshifts. Indeed, their complex nature - which includes the physics of large scale gas cooling, star formation, Active Galactic Nuclei (AGN) feedback, and galaxy mergers and interactions - make them key, if complicated, laboratories for hierarchical processes of structure growth. The observational study of galaxy cluster cores is especially difficult at cosmic noon ($\sim z = 1-2$) where many of the aforementioned processes peak.  Here, the challenge of identifying clusters is coupled with the difficult interpretation of imaging and spectroscopy of high-redshift sources, which have both lower S/N and lower angular resolution.  Nevertheless, progress is being made through wide-field cluster surveys, that are simultaneously able to reveal unique clusters (providing windows onto particular physical  processes) and facilitate global statistical studies of young clusters.

A cluster core at high redshift  contains several important components. Most clusters contain a single large galaxy sitting at the bottom of the gravitational potential well, known as the Brightest Cluster Galaxy (BCG). This object is surrounded by a hot intracluster medium, which, if AGN feedback is lacking or insufficient, may cool onto the BCG \citep{mcnamara2012, mcdonald2019}.  Over time, cluster galaxies are accreted into the core of the cluster, resulting in galaxy-galaxy interactions and mergers with the central BCG. This latter process is further believed (e.g., \cite{1979ApJ...231..659D}) to feed a halo of stars, unbound to any galaxy, around the BCG which themselves produce the Intracluster Light (ICL).

The formation and evolution of BCGs and the ICL, and the role of cooling flows and mergers,  are not well understood.  Semi-analytic models (SAMs) ({\it e.g., } \citet{De_Lucia_2007}; \citet{Tonini_2012}) find that below $z\lesssim$1 the BCG alone grows by a factor of 2-3. 
This appears to be supported by observational evidence \citep{lidman2012},  which indicates that roughly 50\% of tidal remnants from mergers add to the growth of the BCG, whereas the remaining 50\% ends up  in the ICL, though see also \citet{lin2013, oa2014, bellstedt2016}.
Some observational studies, however, have claimed that significant amounts of in-situ star formation occurs within the BCG \citep{webb2015a, mcdonald2016}, while some simulations \citep{puchwein2010} indicate that a significant fraction of the ICL could possibly form in-situ as well.  Indeed, the work of  \citet{hl2020} surmised that the growth of the ICL could be fed by runaway cooling flows, in the absence of AGN heating. The relative importance of in-situ (i.e. the stars formed where we observe them) vs ex-situ (i.e. the stars formed elsewhere, and were accreted via galaxy interactions) star formation for BCGs and the ICL is likely a function of redshift, as is the process driving each (e.g. \citet{mcdonald2016}).   

\subsection{The SpARCS1049 Cluster}\label{subsec:1049}
In this paper, we present the study of the stellar mass build-up in the core of a single system, SpARCSJ-104922.6+564032.5 (hereafter SpARCS1049), located at redshift $z=1.71$. This object  was first detected as a cluster  in the Spitzer Adaptation of the Red-sequence Cluster Survey (SpARCS) \citep{wilson2009, muzzin2009, demarco2010} and published in \citet{Webb_2015}.  This work presented both the cluster itself (through the spectroscopic detection of 27 cluster member galaxies) as well as the high star forming core (860$\pm$130 M$_\odot$yr$^{-1}$) which, at the time, was associated with the BCG. The cluster has a lensing mass of 4$\times$10$^{14}$M$_\odot$ \citep{finner2020} and has been detected in the X-ray by \citet{hl2020}. The latter work, as mentioned in the previous section,  furthermore argued that SpARCS1049 contains a cool core and that the intense star formation is fed by the large scale, runaway cooling of intracluster gas, which has been allowed to cool in the cluster core due to the lack of feedback from the central AGN \citep{trudeau2019}. They surmised that this cooling flow-fed star formation might be forming the ICL in-situ.

Here we present a new analysis of archival Hubble Space Telescope (HST) observations, to assess the current stellar mass present in the region; and the complete analysis of  CO(1-0) observations (briefly shown in \citet{hl2020}), that trace the molecular gas mass of the system. Finally, we add to this previously published infrared data that target the on-going star formation in the core of the cluster. 
Using these three legs we present a census of the past, future and present star formation in the central regions of SpARCS1049 and, we posit, in agreement with \citet{hl2020}, the in-situ build-up of the ICL. 

In \S \ref{sec:observations} we outline the data, including the new Jansky Very Large Array CO(1-0) observations, and we recap the existing Hubble Space Telescope optical imaging and multi-facility infrared imaging and spectroscopy. \S \ref{sec:methods}
outlines our methodology, including the determination of the stellar mass in the vicinity of the BCG and measurement of the molecular gas mass in the same region. We then present our main results in \S \ref{sec:results} and discuss their implications in \S \ref{sec:discussion}. Finally, we conclude and summarize in \S \ref{sec:conclusions}. Throughout the paper we use a standard $\Lambda$ CDM cosmology with H$_\circ$ = 70 km s$^{-1}$ Mpc$^{-1}$; and the kiloparsec to arcsec conversion at the redshift of interest is 1kpc $\sim \: 0.11"$. 

\section{ Observations and Data Reduction} \label{sec:observations}


\subsection{Previous Infrared Imaging and Spectroscopy}
The first detection of the BCG was made by \citet{lonsdale2003} with $Spitzer$-MIPS 24$\mu$m, as part of the SWIRE survey of the Lockman field. As a bright $z\sim 2$ ULIRG candidate, it was followed up by  \citet{farrah2008} with $Spitzer$-IRS which determined its spectroscopic redshift to be $z = 1.7$.  
Finally, \citet{Webb_2015} presented evidence that this object was at the heart of a massive galaxy cluster. This latter paper also presented new SCUBA2 imaging at 450 and 850$\mu$m; and synthesized the infrared Spectral Energy Distribution (SED), including limits and detections derived from the Herschel Space Observatory. Using this extensive data set, they estimated an infrared luminosity of $\sim$ 7 $\times$10$^{12}$ L$_\Sun$ for the BCG, and a corresponding AGN-corrected Star Formation Rate (SFR) \citep{k1998, pope2008} of 860$\pm$130 M$_\Sun$yr$^{-1}$. 

Unfortunately, the PSFs of the infrared instruments are large, with the smallest beam corresponding to 6$^{\prime\prime}$ for  $Spitzer$-MIPS at 24$\mu$m. This made the source of the SFR hard to determine, and lead to associating the emission with the BCG (\citet{farrah2008}). As we will argue in this paper (also in \citet{trudeau2019} and \citet{hl2020}), the location of the bright infrared emission actually lies 1.5$^{\prime\prime}$ ($\sim 13$ kpc) off the BCG, on top of what appears to be galactic debris.



\subsection{Archival Multi-wavelength HST Optical Imaging}
The HST data, first published in \citet{Webb_2015}, were retrieved from the Hubble Legacy Archives, having been targeted by the Hubble Space Telescope's (HST) Wide Field Camera 3 (WFC3) in 2014 and 2015 through Cycle 22 programs\footnote{Based on observations made with the NASA/ESA Hubble
Space Telescope, obtained [from the Data Archive] at the Space
Telescope Science Institute, which is operated by the Association of
Universities for Research in Astronomy, Inc., under NASA contract
NAS 5-26555. These observations are associated with program \#
GO-13677 and \# GO-13747.}.
The object was observed  through the F160W, F105W, and F814W filters with integration times of 9237s, 8543s, and 2846s respectively. 
The data were drizzled into images with 0.06" pixel scale using the STScI software, AstroDrizzle\footnote{http://www.stsci.edu/scientific-community/software/drizzlepac.html}.   

\subsection{New Jansky Very Large Array Observations}\label{subsubsec:radio_img}

We obtained new radio observations using the Jansky Very Large Array (JVLA), in Socorro, New Mexico. We observed SpARCS1049 for a total of 11 hours combining two different observing sessions from January \nth{5} and \nth{12} 2019 (Proposal ID: 18B-177, PI: T. Webb) during optimal observing conditions with an optical depth $\tau_{\SI{42.5}{GHz}}\sim 0.06$. We used the Q-Band in the C-Configuration  which provided a synthesized beam width of $\SI{0.47}{\arcsecond}$.

All of the interferometry data were reduced using the pipeline reduction routine provided by the NRAO center in Socorro. For the CO(1-0) line ($\nu_{rest} = \SI{115.3}{\giga\hertz}$), we expected an observed line frequency of $\sim \SI{42.5}{\giga\hertz}$ considering a previous 
Large Millimetre Telescope redshift measurement of  $z = 1.709$ \citep{webb2017}.

\begin{figure}
    \centering
    \includegraphics[width=\linewidth]{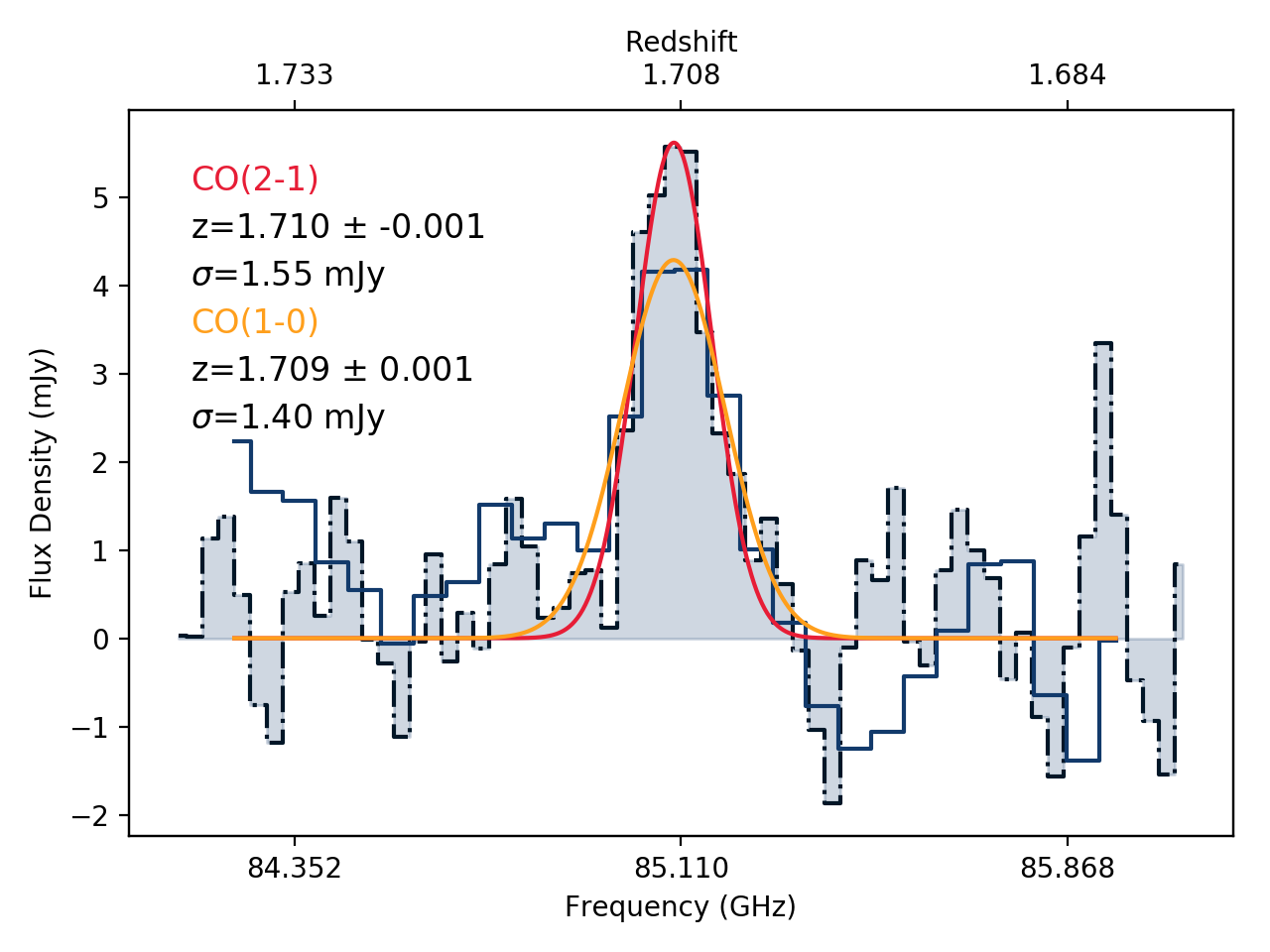}
    \caption{CO line detection from both the 1-0 and previously detected 2-1 \citep{webb2017} transition. The black dotted line represents the CO(2-1) transition line and the solid blue line represents the CO(1-0) transition line. Each have had a Gaussian curve fit to them in order to estimate the redshift. The CO(2-1) line has been scaled in the frequency domain by a factor of 2 to match the CO(1-0) and the CO(2-1) flux density also has been scaled down by a factor of 4 to compensate for the $\nu^2$ SLED effect. \citep{narayanan2014}}.
    \label{fig:spectrum}
\end{figure}

Figure \ref{fig:spectrum} shows the CO(1-0) line detection from the JVLA. The spectrum was extracted in the following iterative manner. First,   using the CASA \textit{tclean} task, we created a cube composed of $\SI{32}{MHz}$-wide slices centered around our target frequency $\nu_{obs} = \SI{42.5}{GHz}$. 
We then visually searched the image slices for emission.  Extended emission was identified at the expected frequency and we then applied an aperture around said emission to extract the spectrum shown in Figure \ref{fig:spectrum}.  

We used a regular Gaussian fitting routine over the previously mentioned channels and measured a peak emission at $42.54781 \pm \SI{0.00009}{\giga\hertz}$, which corresponds to a redshift of $z = 1.708 \pm 0.001$. This agrees well with the previously determined value of $z = 1.709\pm 0.001$ \citep{webb2017}. The FWHM, which corresponds to the measured velocity dispersion of the line, is  $630 \pm \SI{270}{km.s^{-1}}$.    

\begin{figure}
    \centering
    \includegraphics[width=\linewidth]{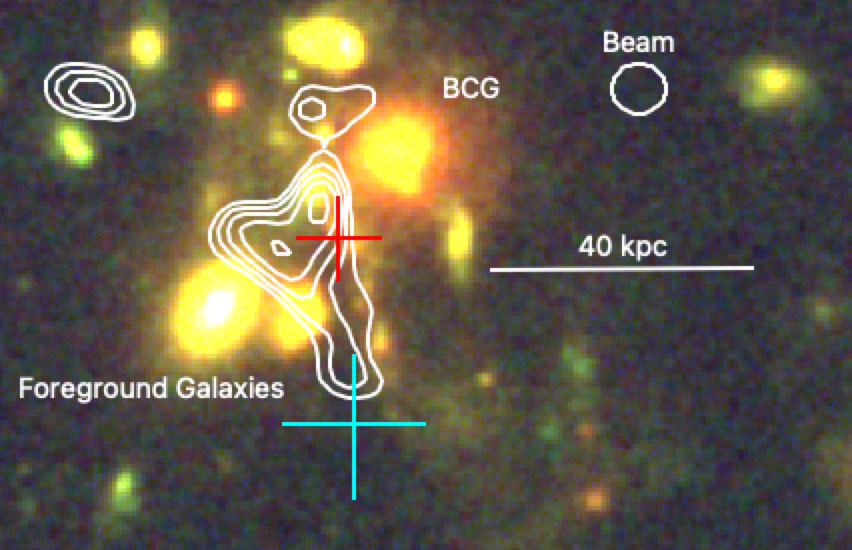}
    \caption{HST optical image of the SpARCS1049 cluster (colour image, using F160W, F140W, and F814W). We identify the BCG and two foreground objects within this map. The white contours show the CO(1-0) map contours starting at 2$\sigma$ and increasing in 0.5$\sigma$ steps. The red cross marks the peak of the MIPS-24$\mu$m detection, and the cyan cross denote the center of the soft X-ray emission. The size of the crosses represent their respective positional uncertainties. The white circle in the upper right corner denotes the radio beam. }
    \label{fig:COMap}
\end{figure}

To create an integrated intensity map of the molecular gas reservoir, we generate a single image from multiple channels.  We combine 2 spectral windows each with 64 channels using CASA's \textit{tclean} task with a resolution of \SI{0.2}{\arcsecond} per pixels for a 1024 pixels square map. We also use a \SI{1}{\arcsecond} taper in \textit{uv}-space such that longer baselines have a much more limited weight on the final image. This usually reduces the overall flux sensitivity, and decreases the resolution. We show this emission overlaid atop the HST optical image in Figure \ref{fig:COMap}. 

\section{Methods and Analysis}\label{sec:methods}
Here we describe the new analyses using the data outlined in the previous section. First we present the extraction of the stellar mass in the region surrounding the BCG, from the HST optical data, followed by a detailed summary of the CO characterization from the JVLA data.
The analysis on the infrared star formation is described thoroughly in \citet{Webb_2015}, where the infrared luminosity was converted to SFR using the \citet{k1998} law, which was then corrected for AGN contribution. No new analysis was done.

\subsection{Measuring the Stellar Mass in the Core}
\subsubsection{Method for Removing Galaxies}\label{subsubsec:galfit}

Imaging observations of clusters poses many challenges due to the large amount of light that blends together from different sources in the cluster and foreground/background.  As we are interested in the central intracluster mass  we need to model and subtract all the nearby galaxies contributing significantly to the light in the region.  We chose the method demonstrated by \citet{Shipley2018} for an iterative processing method to remove galaxies in the \textit{Hubble} Frontier Fields.  We summarize the method here and refer the reader to \citet{Shipley2018} for a detailed discussion.

The procedure measures the isophotal parameters of a galaxy and removes the resulting model as described by \citet{Ferrarese2006}. The modeling method has been shown to work well for elliptical galaxies with significant isophotal twisting, which are the predominant type of bright galaxies in the cluster environment.  The method utilizes the IRAF tasks ELLIPSE to measure the isophotal parameters, which are then passed to the IRAF task BMODEL to create the best fit model for each selected galaxy with statistical finessing of the data to achieve the best fit model even out to large radii (by splining and interpolating the isophotal parameters).  Also, a local background is applied to create a smooth profile at large radii, and the curve of growth is measured to mitigate extra light being modeled that should be attributed to the sky.  Furthermore, the method minimizes effects from poor modeling of a nearby galaxy by masking the image except for the galaxy being modeled.  This is important in the initial stages of modeling in crowded regions.

\begin{figure}
    \centering
    \gridline{\fig{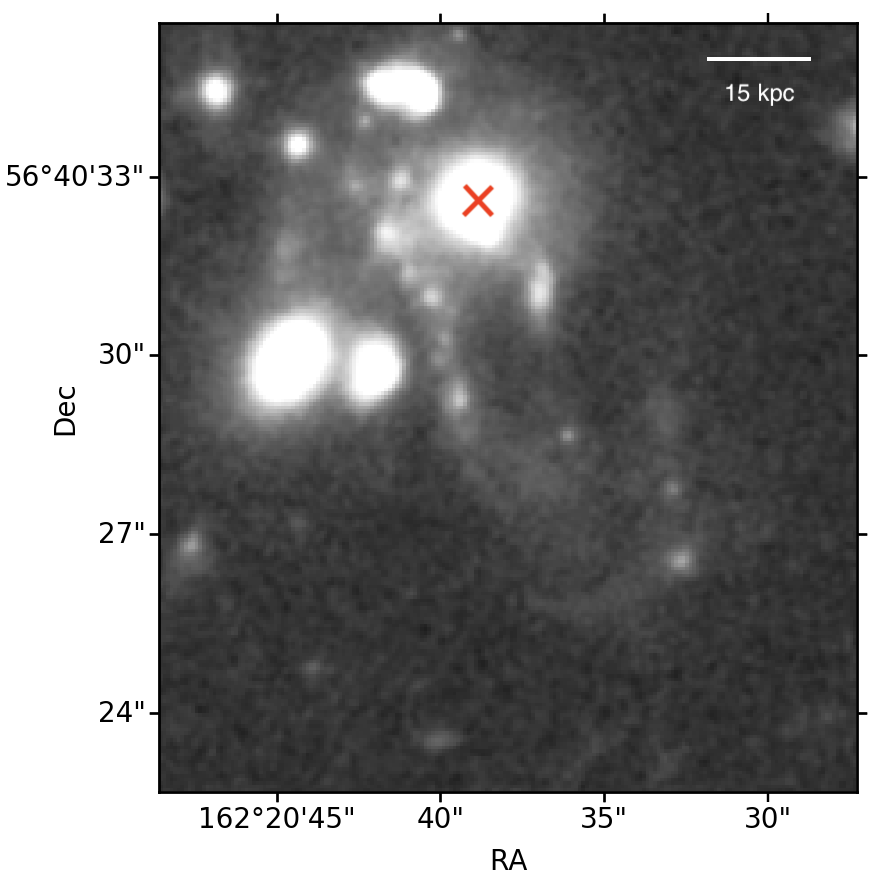}{0.255\textwidth}{(a)}
          \fig{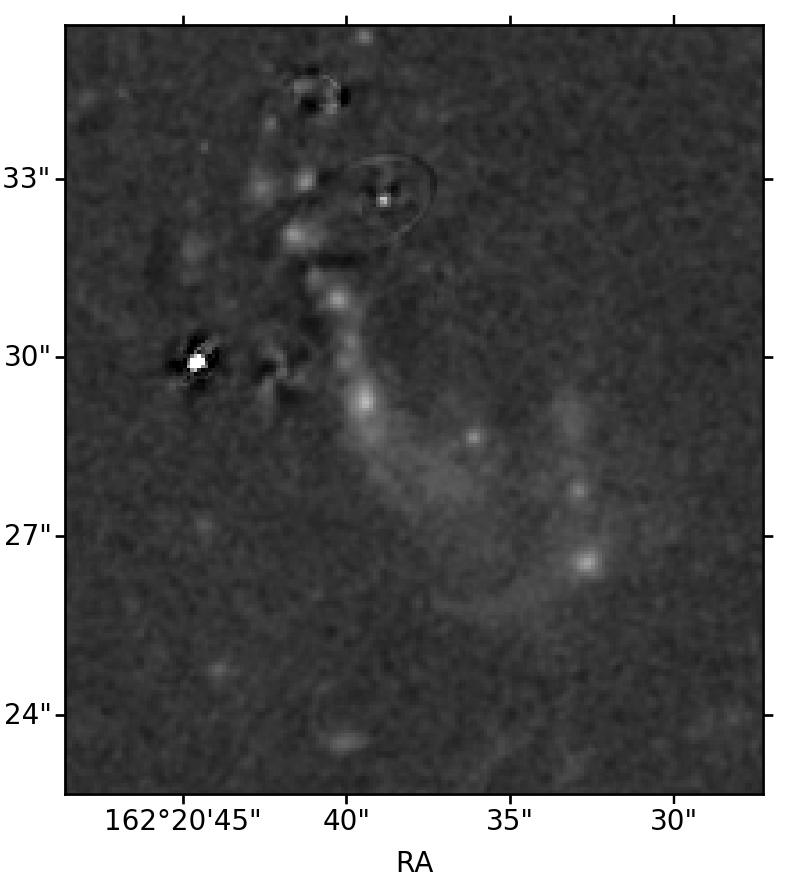}{0.23\textwidth}{(b)}}
    \caption{(a) Original drizzled image of SpARCSJ-104922.6+564032.5 through the F160W filter. The position of the central galaxy is indicated by the red cross. (b) Residual image after the galaxies were fit and removed (see \S \ref{subsubsec:galfit}). }
    \label{fig:1049}
\end{figure}

To achieve improved modeling of the galaxies being removed, the method utilizes an iterative processing step.  After the initial run (described above), the code runs through 10 more iterations of each selected galaxy for 11 total iterations adding in one selected galaxy at a time and modeling it again for each iteration.  The result of this improves the model and the residual for each galaxy selected without having contamination from all other selected galaxies that hindered the initial models.  Furthermore, the method averages over the iterative models to create a better model overall that is more resistant to spurious model results from any one iteration.  The IRAF task IMCOMBINE is used with the following parameters (combine=``average'', reject=``minmax'', nlow=``4'', nhigh=``2'') for the modeled galaxies to achieve this.

We chose to model and subtract 17 galaxies (10 of which are visible on the left panel of Figure \ref{fig:1049}) that were bright, nearby and/or improved the modeling that allows for the best measurement of the intracluster mass near the BCG.  Finally, the residual  patterns left by the subtracted galaxies (primarily in the core of each model) are masked.

\subsection{Estimating the Stellar Mass}\label{sec:stelmass}

Our goal was to derive mass-to-light ratios for each individual pixel of the residual map. To do this, we exploited the relation between optical and near-infrared (NIR) colors and stellar M/L as presented in \cite{2003ApJS..149..289B}, computed using a Salpeter IMF (\citet{salpeter1955}) modified to have only 70\% of the mass.  We used the following equation and parameters:
\begin{equation}
    \log_{10}\left(\frac{M}{L}\right) = a_r + (b_r * color).
    \label{eq:m/l}
\end{equation}
where $a_r = -0.223$ and $b_r = 0.299$ are empirically determined parameters, and the $color$ is the difference in ST magnitudes between the F105W filter and the F160W. Prior to the color estimation, both F105W and F160W magnitudes are k-corrected to map the rest frame magnitudes of Sloan Digital Sky Survey (SDSS) \textit{u} and \textit{r} filters - which are the filters used in determining Equation \ref{eq:m/l} -, as prescribed in \citet{hogg2002}. The k-correction is computed using the \citet{bc03} models, with Chabrier IMF \citep{chabrier} and solar metallicities.

With a mass-to-light ratio in hand we can then convert the observed F160W flux density to a stellar mass, taking into account cosmological dimming. We also measure the mass in the three brightest clumps of the tail, to determine if they have stellar masses comparable to dwarf galaxies.

This method is reproduced twice, once on the data as presented in the right panel of Figure \ref{fig:1049}, and once on a reprocessed version of the data, presented in Figure \ref{fig:1049Rebinned}, re-binned using the \citet{cc2003} 2D implementation of the Voronoi binning technique. We reiterate the mass measurement using the second method as \citet{zibetti2009} estimates that a signal-to-noise ratio (S/N) $\ge$ 20 is necessary to obtain M/L values accurate within 30\%. The repetition of the measurement therefore allows us to obtain a better sense of the variation in the measured stellar mass depending on the methodology.

\subsubsection{Using the Original Data}

\begin{figure}
	\centering
	\includegraphics[width=\linewidth]{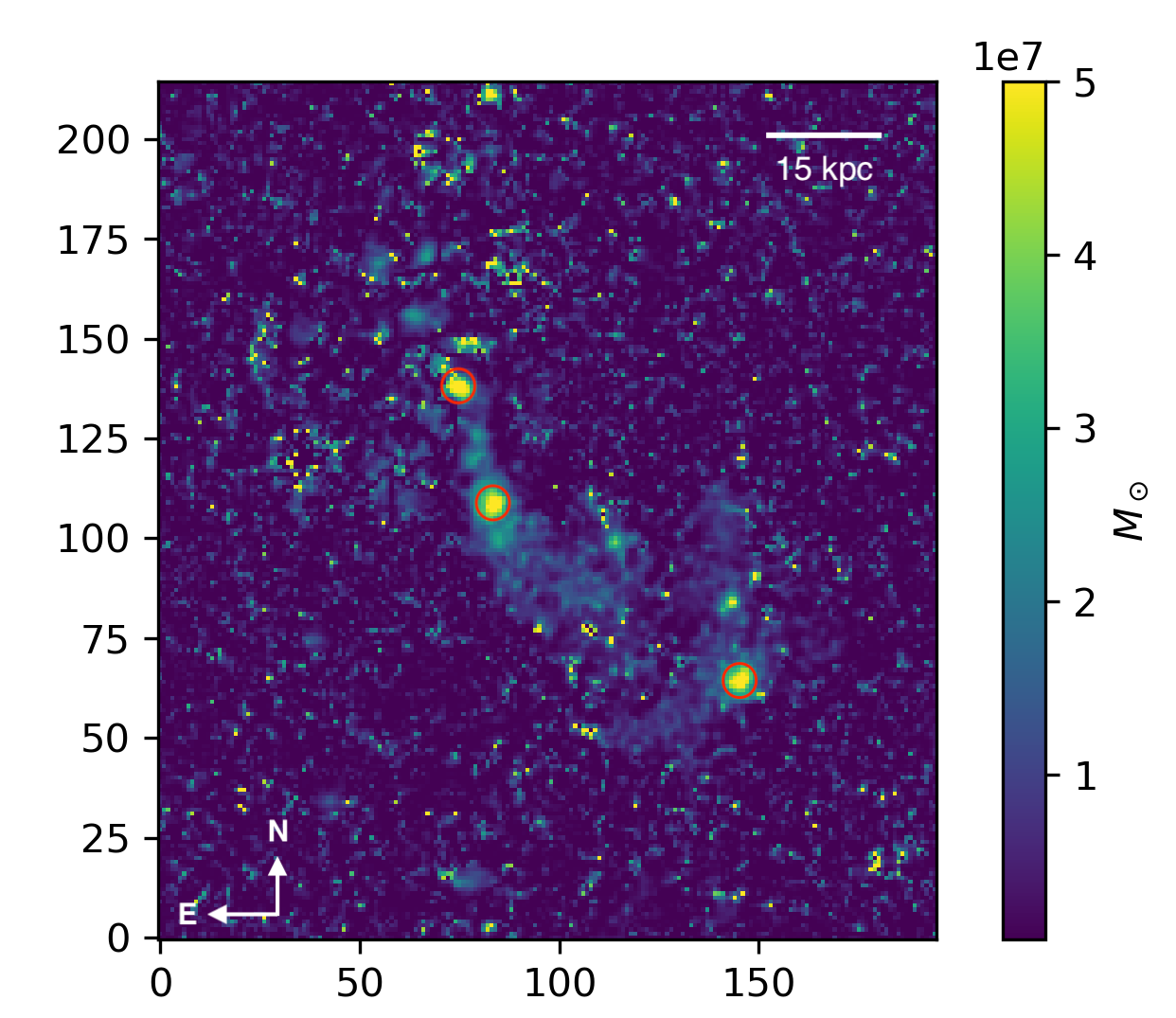}
	\caption{Mass-per-pixel map of the area shown in Figure \ref{fig:1049}. The red apertures circle the three most massive clumps in the region. The noisy pixels in the background are due to pixels with un-physical colours and the galaxy subtraction. These fluctuations are taken into account in the errors budget (see \S \ref{sec:err}). Here we see the mass distribution is compressed into the tail which extends roughly 65 kpc, and is composed of several large mass clumps.}
	\label{fig:mass}
\end{figure}

For the first estimate of the stellar mass, the mass-per-pixel map obtained using the un-binned data is shown in Figure \ref{fig:mass}. This map traces, but is not identical to, the F160W light map; in which some of the clumps are more defined. \par 
We note that sky subtraction - one step of data reduction through AstroDrizzle - returned negative values for some of the background pixels in both HST images. This leads to errors when converting the data to magnitudes before getting the color. To overcome this issue, the steps described in the paragraph above were repeated twice, once on the absolute value of all the pixels in the image, and once on the absolute value of the pixels flagged as negative. The final mass value is then the difference between the sum of all the pixels in the mass pixel map (Figure \ref{fig:mass}) and the sum of the``negative" pixels.


In addition, noise in the colour values lead to high M/L values for some pixels, which then transferred to high mass values. To estimate the magnitude of this effect, we set a threshold value above which the M/L of the pixel is deemed un-physical, and replaced by a background value. By varying the value of the threshold from the lowest (using the minimum threshold below which physical mass in the tail is cut-off) to infinity (without a threshold), we observe that the mass varies around $6.6 \times 10^{10} \: M_\odot$, with a 10\% scatter on this value, which is used as the uncertainty on the mass from this effect. It should be noted that Figure \ref{fig:mass} displays the pixel map of the mass for a threshold value of 10.

Finally, despite the sky subtraction step in AstroDrizzle, we included background fluctuations in our error analysis. To do so, we divide the area around the tail of SpARCSJ-1049 into 64 squares of 20x20 pixels; in which we replace the pixel values associated to light sources by randomly chosen sky values. We proceed to compute the mean values of the pixels in each of the squares, and use the standard deviation of those 64 values as the uncertainty in the pixel values. This uncertainty can then be propagated to both the color and luminosity, and finally to the mass, using error propagation formulas. This amounts to an error of $<1\%$ in the final mass measurement.

\subsubsection{Using the Binned Data}

\begin{figure}
    \centering
       \gridline{\fig{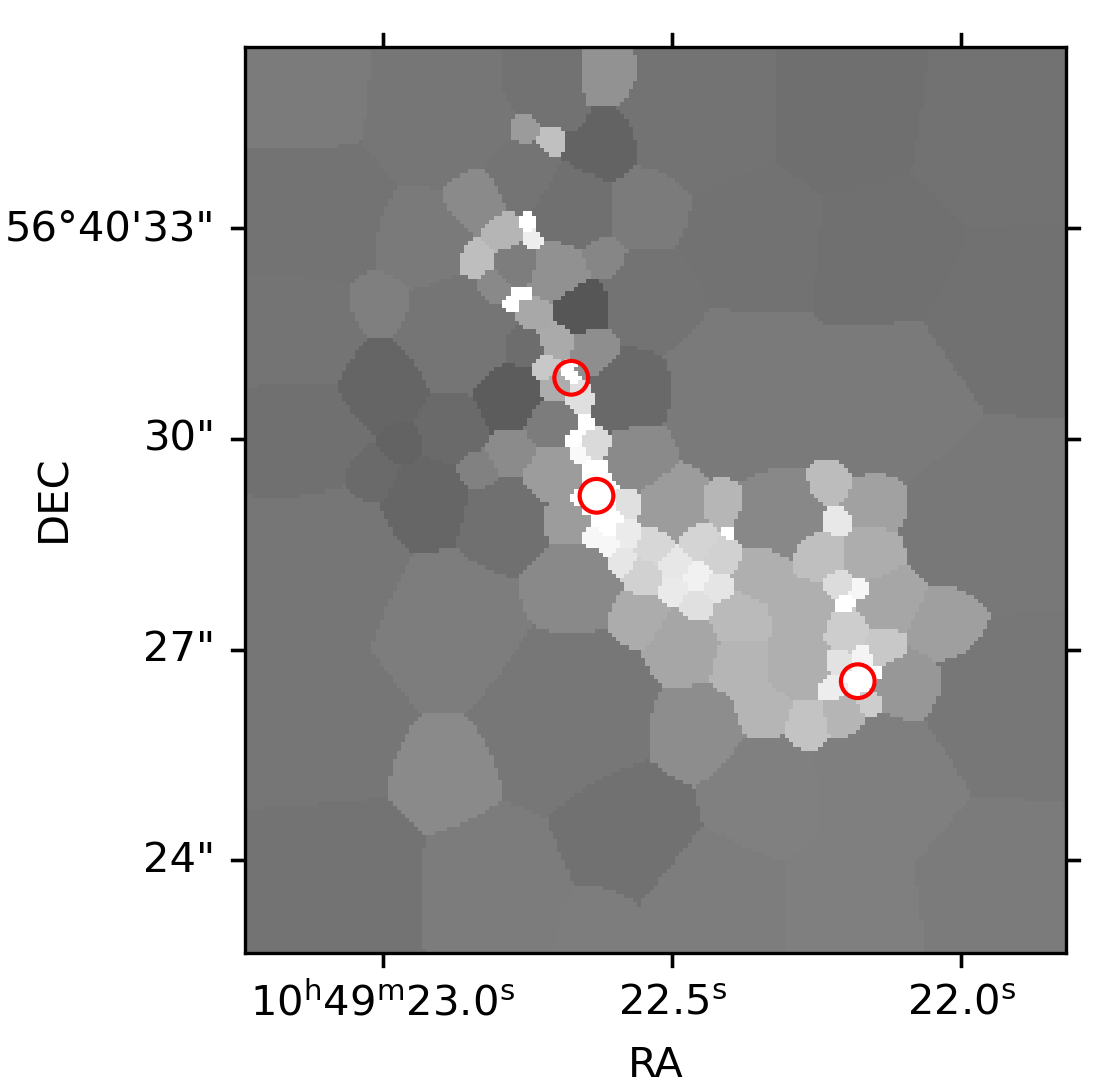}{0.48\linewidth}{(a)}
          \fig{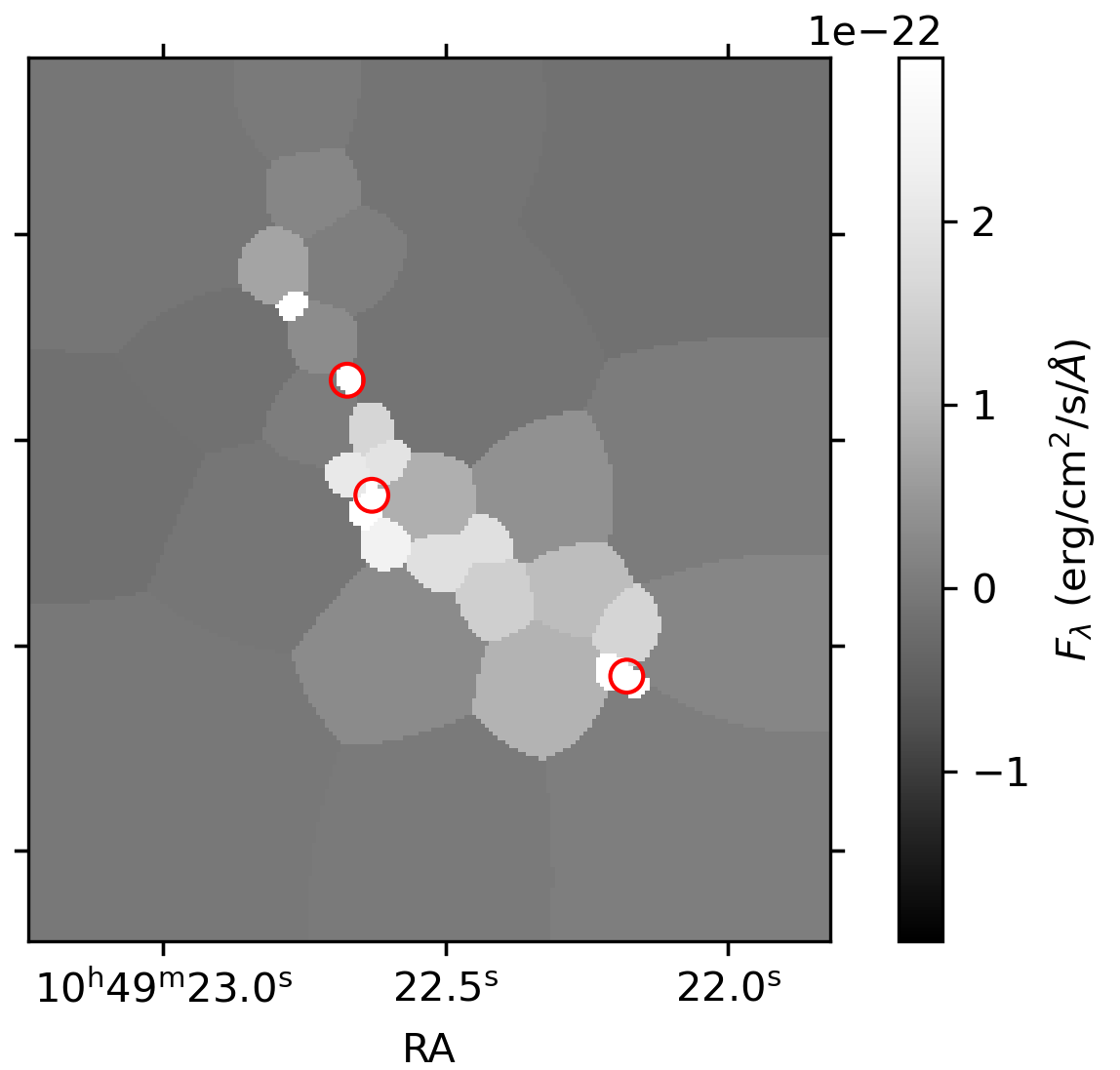}{0.49\linewidth}{(b)}}
    \caption{Resulting F160W image after reprocessing the data using the Voronoi binning, showing the average flux density value per bin, where (a) had a target S/N of 10, and (b) had a target S/N of 20. The mass measurements were made using the S/N = 20 images, but we note that both returned values in agreement with each other. The red annuli circle the same clumps as in Figure \ref{fig:mass}.}
    \label{fig:1049Rebinned}
\end{figure}

For the second estimate of the stellar mass, both the F160W and F105W images are re-processed using the 2D implementation of the Voronoi binning technique by \citet{cc2003}. The results are shown in Figure \ref{fig:1049Rebinned} for the F160W data, which displays the average flux density per bin using two different target S/N. While the number of bins varies with the S/N, the overall shape of the structure is well preserved. We also re-process the F105W data using the binning map of the F160W image.

We then proceeded to reproduce the analysis presented in \S \ref{sec:stelmass} on the new data, to obtain another estimate of the stellar mass. Propagating the errors from each bin to the mass only adds a 1\% uncertainty on the mass measurement.

\subsubsection{Additional Uncertainties in the Mass Estimate}\label{sec:err}

There are several uncertainties inherent or introduced in both of our measurements by the method outlined in \S \ref{sec:stelmass}.

\begin{itemize}

\item Varying the input model used to compute the k-correction adds an additional 4\% error in the mass measurement.

\item Other studies have found varying values for $a_\lambda$ and $b_\lambda$ in Equation \ref{eq:m/l}. We repeated the mass measurement using values from \citet{zibetti2009} and \citet{gb2019}, and found varying the parameters introduces a 14\% error in the stellar mass from the original data, and a 20\% error in the stellar mass from the binned data.

\end{itemize}



\subsection{Estimating the Molecular Gas Mass}\label{subsubsec:radio}

We now discuss the methodology of measuring the molecular gas mass using the JVLA radio data.
We use the spectral tool from the \textit{casaviewer} on the cube image mentioned in \S \ref{subsubsec:radio_img} to generate a data file containing the spectrum of a 6$^{\prime\prime} \times $6$^{\prime\prime}$  circular aperture around our source peak.

To estimate the flux in the line, with an associated error we adopt a Monte Carlo approach. We generate 20 000 new realizations of the line using the rms uncertainties in each channel. We can then fit and integrate a Gaussian curve on each of these generated emission lines and compute the mean and standard deviation of the integrated line flux: $S_{\text{CO}}\Delta\nu = 1.0\pm \SI{0.6}{\jansky\kilo\meter\per\second}$.


We compute the CO integrated source brightness temperature using equation \ref{eq:LCO}, where $D_L$ is the luminosity distance, $z$ is the redshift, $\nu_{obs}$ is the observed frequency mentioned above.  

\begin{equation}
    L_{\text{CO}}' = \SI{3.25e7}\times\ S_{\text{CO}}\Delta\nu \frac{D_L^2}{(1+z)^3 \nu_{obs}^2}\si{\kelvin\kilo\meter\per\second\parsec\squared}
    \label{eq:LCO}
\end{equation}

We find  $L_{\text{CO}}' =  (1.3\pm 0.9) \times 10^{11} \si{\kelvin \kilo \meter \per \second \parsec \squared}$.

The integrated source brightness temperature can be converted into a \ce{H_2} molecular gas mass using a conversion factor $\alpha_{\text{CO}}$. Here we employ $\alpha_{\text{CO}} = 0.8$, which is typical for ultra-luminous infrared galaxies (ULIRG) and also allows for a conservative mass estimate \citep{webb2017}.  Using equation \ref{eq:MH2} we compute a mass $M_{\ce{H_2}} = 1.0\pm\SI{0.7e11}{\sunM}$. 

\begin{equation}
    \frac{M_{\ce{H_2}}}{\solar{M}} = \alpha_{\text{CO}}\frac{L_{\text{CO}}'}{\si{\kelvin\kilo\meter\per\second\parsec\squared}
    }
    \label{eq:MH2}
\end{equation}

\section{Results}\label{sec:results}
In this paper we set out to synthesize three important measurements: (1) the stellar mass in the vicinity of the BCG (but not within the BCG proper) -  representing the history of the star formation within the core; (2) the molecular gas mass around the BCG - representing the future star formation within the core; and (3) the total current star formation rate occurring on or near the BCG.  We now discuss each in turn.

The total stellar mass in the $12.9''\times 11.7''$ cutout is measured using the methods described in \S \ref{sec:stelmass}, and is dominated by the mass within the clumpy tidal tail.
This returns final values for the stellar mass in the BCG's vicinity of  $M_{near} = (6.6 \pm 1.2) \times 10^{10} \: M_\odot$ for the original data and $M_{near} = (2.2 \pm 0.5) \times 10^{10} \: M_\odot$ for the re-binned data. We note that the BCG-proper has previously been estimated to contain (3.0$\pm0.4)\times$10$^{11}$ M$_\odot$ \citep{Webb_2015}. The totals of these measurements ($(3.7\pm 0.4)
\times10^{11}M_\odot$ and $(3.2\pm 0.4)
\times10^{11}M_\odot$) sit below the measurement of the mass of the ICL (within 100kpc radius)+BCG undertaken by \citet{demaio2020} for SpARCS1049 of (5$\pm$0.5)$\times$10$^{11}$M$_\odot$.

The three brightest clumps in the tail (circled in red, Figure \ref{fig:mass}) have stellar masses (from top to bottom) $1.1 \times 10^{9} \: M_\odot$, $1.3 \times 10^{9} \: M_\odot$, and $1.2 \times 10^{9} \: M_\odot$; which indicates that they might be dwarf galaxies formed by the intense star formation. They only amount to $\sim 6\%$ of the total mass measured in the area.

The molecular gas mass is measured in \S \ref{subsubsec:radio} to be 1.0$\pm$0.7$\times$10$^{11}$M$_\odot$.  The high spatial resolution of the JVLA clearly spatially isolates this mass to an area roughly 12kpc distant from the BCG, and coincident with the clumpy tail that contains the bulk of the stellar mass.  This mass is in agreement with the previous large beam (23$^{\prime\prime}$) CO(2-1) mass estimate of \citet{webb2017} of (1.1$\pm$0.1)$\times$10$^{11}$M$_\odot$.  The good agreement between these two measurements indicates all of the molecular gas is located within the 12 kpc (radius) clump resolved by the JVLA, and we are not missing extended flux in the interferometric measurement.  

Finally, we reiterate the measurement of the star formation rate in a large - $\sim$ 50 kpc -  aperture (that is, with a 6$^{\prime\prime}$ beam). The positional peak of this flux is coincident with the peak of the molecular gas reservoir.  After removing flux attributed to a weak AGN \citep{pope2008}, \citet{Webb_2015} estimate a star formation rate of SFR = 860$\pm$130M$_\odot$yr$^{-1}$. To place this value in the same context as the two above masses we note that for a duty cycle of 1Gyr this star formation would produce $\sim$ 9$\times$10$^{11}$M$_\odot$.

\section{Discussion}\label{sec:discussion}
The results summarized in the previous section indicate that a large amount of stellar mass is  being assembled in the core (inner $\sim$100 kpc) of SpARCS1049. Building on the work from \citet{Webb_2015} and \citet{hl2020} - the latter of which surmised that the cooling-flow was feeding the \textit{in-situ} formation of the ICL - these results add to the picture that a significant amount of mass has already been built up and the fuel is in place to continue to feed the star formation for some time ($\sim$1Gyr). Moreover, the HST and JVLA imaging indicate that this mass is not being formed within the BCG itself, but rather off-axis by $\sim$ 12 kpc and extended along a long filamentary structure that stretches for $\sim$65 kpc. Although no dynamical  information is yet available it is unlikely, given the projected distance (tens of kpc) that this material is bound to the BCG and we instead suggest that it will be incorporated into the ICL. 


Intergalactic star formation in intracluster  space has been reported in many low-redshift systems \citep{sun2007, sun2010, lee2000, gerhard2002, cortese2004, cortese2007, yoshida2008, smith2010, sivan2010, gull2020}, and explored by simulations \citep{ss2001, vollmer2001, kronberger2008, kapferer2009, puchwein2010}. This star formation occurs within long galactic tails of material that has been stripped out of infalling galaxies and is generally seen to be low levels of activity (c.f. \citet{cortese2007} who report LIRG-levels of star formation of  $\sim$50 M$_\odot$yr$^{-1}$), contrasting with the very high level of SF activity seen here. Indeed, the large star formation rate in  the centre of SpARCS1049 is reminiscant of the excessive cooling seen in many hydrodynamical simulations without AGN feedback (e.g., \citet{puchwein2008, puchwein2010, sijacki2007, mccarthy2010}). These unhindered cooling flows over-produce the stars in BCGs and make the stellar population too blue. Most current hydrodynamical simulations thus assume strong AGN feedback and disallow intergalactic star formation. 

\citet{hl2020} however, showed that the SpARCS1049 cluster contains a cool-core and suggested that the star formation was being fueled by just such a large-scale runaway cooling flow of the hot intracluster X-ray gas. This was based on deep Chandra X-ray observations, which shows that the coolest intracluster X-ray gas (0.7-1.0 keV) was remarkably co-spatial with the stellar mass and molecular gas that we see here. \citet{hl2020} further conjectured that the sloshing off-axis of the BCG (which contains a weak AGN, \citet{trudeau2019}) has allowed this cooling to occur in the absence of central AGN heating. We expect this combination to support the star formation already present in the cluster, and transform most of the molecular gas reservoir into intra-cluster stars. While it is possible that feedback from supernovae blows away part of the gas, recent studies indicate that it cannot by itself quench star formation in high-mass systems \citep{henriques2019,bluck2019}, and without feedback from the AGN; it is unlikely that the SF in this clusters stops before running out of gas.

Coupled with the simulations this would suggest that a dynamically unrelaxed cluster with suppressed AGN heating is necessary for such in-situ ICL growth and that the duty cycle for such an event would be roughly the relaxation time of the sloshed BCG. This chain of events is similar to that of a $z=0.485$ cluster observed in the Clusters Hiding in Plain Sight (CHiPS) survey \citep{Somboonpanyakul2021}; in which galaxy interactions (in this case a major merger) triggered off-axis cooling of low-entropy gas, itself leading to active star formation.

Exactly what fraction of the ICL is forming is difficult to assess, given the uncertain constraints on the total mass of the ICL at these redshifts. Observational measurements \citep{furnell2021, gonzalez2005, gonzalez2010, sand2011, gal-yam2003, zibetti2005, kb2007} indicate that for a halo of the mass of SpARCS1049 (M$_{200}$ = 4$\times$10$^{14}$M$_\odot$) approximately 10-20\% (inside roughy r$_{200}$) of the stellar mass will be contained in the ICL at low redshift.  If the mass of the ICL and cluster galaxies co-evolves then we can apply this ratio to high redshift.  A back of the envelope calculation can thus be done: SpARCS1049  would have stellar mass in galaxies of roughly 8$\times$10$^{12}$ (using $f_{500}^{stars} \sim 0.02$, \citet{giodini2009}), and thus the expected ICL mass would be $\sim$8$\times$10$^{11}$ M$_\odot$. Given that we see $\sim$2.2$\times$10$^{10}$M$_\odot$ to $\sim$6.6$\times$10$^{10}$M$_\odot$ of stellar mass in place, and another $\sim$1$\times$10$^{11}$M$_\odot$ of fuel, with ample star formation to use up this fuel in a short amount of time, SpARCS1049 will have amassed a significant portion ($\sim 15$ to $21$\%, depending on the data processing method) of its ICL through this mechanism.
Such an event can be roughly encompassed by the constraints on the ICL placed by studies of the Virgo cluster.  \citet{williams2007} find that 20-30\% of the intracluster stellar population in Virgo is younger than 10Gyr and conclude that Virgo's intracluster stars must have multiple origin mechanisms.   Indeed, \citet{edwards2020} estimate an average age of the ICL of 9.2$\pm$3.5 Gyr.

This is in sharp contrast to what is prescribed in semi-analytical models however, where the ICL forms through the stripped stellar mass  of cluster galaxies, during interactions with the BCG or between themselves (e.g., \citet{contini2018}, see also \citet{contini2021} for a review of the ICL).  In these simulations, AGN feedback is  invoked to prohibit excessive cooling, and intergalactic 
star formation is, in general,  not allowed to occur. Thus, the lack of any \textit{in-situ} star formation is not a direct prediction of the models, but is written into the physics.

We note also, that a recent publication has presented an alternative view of the SpARCS1049 core.
\citet{ccs2020} suggest that the immense amount of gas seen in the central region of the cluster was delivered, not by a cooling flow, but through gas clouds stripped from an infalling galaxy as seen in many clusters at lower redshift. This distant single Jelly Fish galaxy is destined, in their view, to merge with the BCG and contribute to its stellar mass growth. However, this explanation for the cluster is unlikely, as the X-ray emission is too massive and bright to be attributed to a jellyfish galaxy; and can only be associated with a massive cool core. Moreover, the fact that the cool 0.7-1.0 keV gas and the optical tail are remarkably co-spatial makes it all the more probable that the two emissions are related. Further observations with better spatial resolution would be needed to further confirm this postulate, in particular surrounding the star forming region. Additional information regarding the dynamics at play in that same region would also allow to distinguish between the two hypotheses mentioned in this discussion. 


\section{Conclusions}\label{sec:conclusions}

Here we present and summarize evidence that, taken together, suggests that we are witnessing the in-situ buildup a substantial amount of stellar mass in the core of SpARCS1049. A summary of our specific conclusions is as follows:

\begin{itemize}
    \item We present two estimates of the sum of the established stellar mass in the clumpy tidal tail within the core of the SpARCS1049 cluster, one made using the original HST imaging, and one made using re-processed data. We find that, over a  $12.9''\times 11.7''$ region, but dominated by the tail, there exist a substantial amount of stellar mass: $(6.6 \pm 1.2) \times 10^{10} \: M_\odot$ using the first method, and $(2.2 \pm 0.5) \times 10^{10} \: M_\odot$ using the second one. The scatter in Equation \ref{eq:m/l} also introduces an average systematic error of $\sim 40\%$ \citep{2003ApJS..149..289B}.

    
    \item{We present a new CO(1-0) measurement of the total molecular gas mass within a single large clump in the core of SpARCS1049.  We measure this to be 1.0$\pm$0.7$\times$10$^{11}$M$_\odot$. }
    
    \item{Both the large amount of stellar mass and the vast reservoir of molecular gas lie roughly 12 kpc off the BCG.  We therefore suggest, in line with \cite{hl2020}, that we are not seeing the BCG stellar mass assembly, but rather the in-situ growth of the ICL.}
    
    \item{Simple arguments indicate that if this is indeed the in-situ growth of the ICL, it may be forming a significant fraction ($\sim 15$ to $21$\%) of the ICL in this cluster.}
    

\end{itemize}

\facilities{HST(WFC3), JVLA}
\software{Astropy \citep{2013A&A...558A..33A},  
          CASA \citep{casa}, 
          SExtractor \citep{1996A&AS..117..393B}
          }

\acknowledgments
C.B. and T.M.A.W. acknowledges support from NSERC via the Discovery and from the Fonds de Recherche du Qu\'ebec - Nature et Technologies (FRQNT).  
J.H.-L. acknowledges support from NSERC via the Discovery grant program, the Accelerator grant program, as well as the Canada Research Chair program. GW gratefully acknowledges support from the National Science Foundation through grant AST-1517863, from HST program number GO-15294, and from grant number 80NSSC17K0019 issued through the NASA Astrophysics Data Analysis Program (ADAP). Support for program number GO-15294 was provided by NASA through a grant from the Space Telescope Science Institute, which is operated by the Association of Universities for Research in Astronomy, Incorporated, under NASA contract NAS5-26555. We thank Vivian Tan and Visal Sok for their help with the Voronoi binning. We thank the JVLA staff with their immense help reducing and analysing the data. The National Radio Astronomy Observatory is a facility of the National Science Foundation operated under cooperative agreement by Associated Universities, Inc. This research is based on observations made with the NASA/ESA Hubble Space Telescope obtained from the Space Telescope Science Institute, which is operated by the Association of Universities for Research in Astronomy, Inc., under NASA contract NAS 5–26555. These observations are associated with program(s) \# GO-13677 and \# GO-13747. \par

%

%





\bibliography{sample63.bib}{}

\begin{thebibliography}{}
\expandafter\ifx\csname natexlab\endcsname\relax\def\natexlab#1{#1}\fi
\providecommand{\url}[1]{\href{#1}{#1}}
\providecommand{\dodoi}[1]{doi:~\href{http://doi.org/#1}{\nolinkurl{#1}}}
\providecommand{\doeprint}[1]{\href{http://ascl.net/#1}{\nolinkurl{http://ascl.net/#1}}}
\providecommand{\doarXiv}[1]{\href{https://arxiv.org/abs/#1}{\nolinkurl{https://arxiv.org/abs/#1}}}

\bibitem[{{Astropy Collaboration} {et~al.}(2013){Astropy Collaboration},
  {Robitaille}, {Tollerud}, {Greenfield}, {Droettboom}, {Bray}, {Aldcroft},
  {Davis}, {Ginsburg}, {Price-Whelan}, {Kerzendorf}, {Conley}, {Crighton},
  {Barbary}, {Muna}, {Ferguson}, {Grollier}, {Parikh}, {Nair}, {Unther},
  {Deil}, {Woillez}, {Conseil}, {Kramer}, {Turner}, {Singer}, {Fox}, {Weaver},
  {Zabalza}, {Edwards}, {Azalee Bostroem}, {Burke}, {Casey}, {Crawford},
  {Dencheva}, {Ely}, {Jenness}, {Labrie}, {Lim}, {Pierfederici}, {Pontzen},
  {Ptak}, {Refsdal}, {Servillat}, \& {Streicher}}]{2013A&A...558A..33A}
{Astropy Collaboration}, {Robitaille}, T.~P., {Tollerud}, E.~J., {et~al.} 2013,
  \aap, 558, A33, \dodoi{10.1051/0004-6361/201322068}

\bibitem[{{Bell} {et~al.}(2003){Bell}, {McIntosh}, {Katz}, \&
  {Weinberg}}]{2003ApJS..149..289B}
{Bell}, E.~F., {McIntosh}, D.~H., {Katz}, N., \& {Weinberg}, M.~D. 2003, \apjs,
  149, 289, \dodoi{10.1086/378847}

\bibitem[{{Bellstedt} {et~al.}(2016){Bellstedt}, {Lidman}, {Muzzin}, {Franx},
  {Guatelli}, {Hill}, {Hoekstra}, {Kurinsky}, {Labbe}, {Marchesini}, {Marsan},
  {Safavi-Naeini}, {Sif{\'o}n}, {Stefanon}, {van de Sande}, {van Dokkum}, \&
  {Weigel}}]{bellstedt2016}
{Bellstedt}, S., {Lidman}, C., {Muzzin}, A., {et~al.} 2016, \mnras, 460, 2862,
  \dodoi{10.1093/mnras/stw1184}

\bibitem[{{Bertin} \& {Arnouts}(1996)}]{1996A&AS..117..393B}
{Bertin}, E., \& {Arnouts}, S. 1996, \aaps, 117, 393,
  \dodoi{10.1051/aas:1996164}

\bibitem[{{Bluck} {et~al.}(2019){Bluck}, {Bottrell}, {Teimoorinia},
  {Henriques}, {Mendel}, {Ellison}, {Thanjavur}, {Simard}, {Patton},
  {Conselice}, {Moreno}, \& {Woo}}]{bluck2019}
{Bluck}, A. F.~L., {Bottrell}, C., {Teimoorinia}, H., {et~al.} 2019, \mnras,
  485, 666, \dodoi{10.1093/mnras/stz363}

\bibitem[{{Bruzual} \& {Charlot}(2003)}]{bc03}
{Bruzual}, G., \& {Charlot}, S. 2003, \mnras, 344, 1000,
  \dodoi{10.1046/j.1365-8711.2003.06897.x}

\bibitem[{{Cappellari} \& {Copin}(2003)}]{cc2003}
{Cappellari}, M., \& {Copin}, Y. 2003, \mnras, 342, 345,
  \dodoi{10.1046/j.1365-8711.2003.06541.x}

\bibitem[{{Castignani} {et~al.}(2020){Castignani}, {Combes}, \&
  {Salom{\'e}}}]{ccs2020}
{Castignani}, G., {Combes}, F., \& {Salom{\'e}}, P. 2020, \aap, 635, L10,
  \dodoi{10.1051/0004-6361/201937155}

\bibitem[{{Chabrier}(2003)}]{chabrier}
{Chabrier}, G. 2003, \pasp, 115, 763, \dodoi{10.1086/376392}

\bibitem[{{Contini}(2021)}]{contini2021}
{Contini}, E. 2021, arXiv e-prints, arXiv:2107.04180.
\newblock \doarXiv{2107.04180}

\bibitem[{{Contini} {et~al.}(2018){Contini}, {Yi}, \& {Kang}}]{contini2018}
{Contini}, E., {Yi}, S.~K., \& {Kang}, X. 2018, \mnras, 479, 932,
  \dodoi{10.1093/mnras/sty1518}

\bibitem[{{Cortese} {et~al.}(2004){Cortese}, {Gavazzi}, {Boselli}, \&
  {Iglesias-Paramo}}]{cortese2004}
{Cortese}, L., {Gavazzi}, G., {Boselli}, A., \& {Iglesias-Paramo}, J. 2004,
  \aap, 416, 119, \dodoi{10.1051/0004-6361:20031755}

\bibitem[{{Cortese} {et~al.}(2007){Cortese}, {Marcillac}, {Richard},
  {Bravo-Alfaro}, {Kneib}, {Rieke}, {Covone}, {Egami}, {Rigby}, {Czoske}, \&
  {Davies}}]{cortese2007}
{Cortese}, L., {Marcillac}, D., {Richard}, J., {et~al.} 2007, \mnras, 376, 157,
  \dodoi{10.1111/j.1365-2966.2006.11369.x}

\bibitem[{De~Lucia \& Blaizot(2007)}]{De_Lucia_2007}
De~Lucia, G., \& Blaizot, J. 2007, Monthly Notices of the Royal Astronomical
  Society, 375, 2–14, \dodoi{10.1111/j.1365-2966.2006.11287.x}

\bibitem[{{DeMaio} {et~al.}(2020){DeMaio}, {Gonzalez}, {Zabludoff}, {Zaritsky},
  {Aldering}, {Brodwin}, {Connor}, {Donahue}, {Hayden}, {Mulchaey},
  {Perlmutter}, \& {Stanford}}]{demaio2020}
{DeMaio}, T., {Gonzalez}, A.~H., {Zabludoff}, A., {et~al.} 2020, \mnras, 491,
  3751, \dodoi{10.1093/mnras/stz3236}

\bibitem[{{Demarco} {et~al.}(2010){Demarco}, {Wilson}, {Muzzin}, {Lacy},
  {Surace}, {Yee}, {Hoekstra}, {Blindert}, \& {Gilbank}}]{demarco2010}
{Demarco}, R., {Wilson}, G., {Muzzin}, A., {et~al.} 2010, \apj, 711, 1185,
  \dodoi{10.1088/0004-637X/711/2/1185}

\bibitem[{{Dressler}(1979)}]{1979ApJ...231..659D}
{Dressler}, A. 1979, \apj, 231, 659, \dodoi{10.1086/157229}

\bibitem[{{Edwards} {et~al.}(2020){Edwards}, {Salinas}, {Stanley}, {Holguin
  West}, {Trierweiler}, {Alpert}, {Coelho}, {Koppaka}, {Tremblay}, {Martel}, \&
  {Li}}]{edwards2020}
{Edwards}, L. O.~V., {Salinas}, M., {Stanley}, S., {et~al.} 2020, \mnras, 491,
  2617, \dodoi{10.1093/mnras/stz2706}

\bibitem[{{Farrah} {et~al.}(2008){Farrah}, {Lonsdale}, {Weedman}, {Spoon},
  {Rowan-Robinson}, {Polletta}, {Oliver}, {Houck}, \& {Smith}}]{farrah2008}
{Farrah}, D., {Lonsdale}, C.~J., {Weedman}, D.~W., {et~al.} 2008, \apj, 677,
  957, \dodoi{10.1086/529485}

\bibitem[{{Ferrarese} {et~al.}(2006){Ferrarese}, {C{\^o}t{\'e}}, {Jord{\'a}n},
  {Peng}, {Blakeslee}, {Piatek}, {Mei}, {Merritt}, {Milosavljevi{\'c}},
  {Tonry}, \& {West}}]{Ferrarese2006}
{Ferrarese}, L., {C{\^o}t{\'e}}, P., {Jord{\'a}n}, A., {et~al.} 2006, \apjs,
  164, 334, \dodoi{10.1086/501350}

\bibitem[{{Finner} {et~al.}(2020){Finner}, {James Jee}, {Webb}, {Wilson},
  {Perlmutter}, {Muzzin}, \& {Hlavacek-Larrondo}}]{finner2020}
{Finner}, K., {James Jee}, M., {Webb}, T., {et~al.} 2020, \apj, 893, 10,
  \dodoi{10.3847/1538-4357/ab7bdb}

\bibitem[{{Furnell} {et~al.}(2021){Furnell}, {Collins}, {Kelvin}, {Baldry},
  {James}, {Manolopoulou}, {Mann}, {Giles}, {Bermeo}, {Hilton}, {Wilkinson},
  {Romer}, {Vergara}, {Bhargava}, {Stott}, {Mayers}, \& {Viana}}]{furnell2021}
{Furnell}, K.~E., {Collins}, C.~A., {Kelvin}, L.~S., {et~al.} 2021, \mnras,
  502, 2419, \dodoi{10.1093/mnras/stab065}

\bibitem[{{Gal-Yam} {et~al.}(2003){Gal-Yam}, {Maoz}, {Guhathakurta}, \&
  {Filippenko}}]{gal-yam2003}
{Gal-Yam}, A., {Maoz}, D., {Guhathakurta}, P., \& {Filippenko}, A.~V. 2003,
  \aj, 125, 1087, \dodoi{10.1086/346141}

\bibitem[{{Garc{\'\i}a-Benito} {et~al.}(2019){Garc{\'\i}a-Benito},
  {Gonz{\'a}lez Delgado}, {P{\'e}rez}, {Cid Fernandes}, {S{\'a}nchez}, \& {de
  Amorim}}]{gb2019}
{Garc{\'\i}a-Benito}, R., {Gonz{\'a}lez Delgado}, R.~M., {P{\'e}rez}, E.,
  {et~al.} 2019, \aap, 621, A120, \dodoi{10.1051/0004-6361/201833993}

\bibitem[{{Gerhard} {et~al.}(2002){Gerhard}, {Arnaboldi}, {Freeman}, \&
  {Okamura}}]{gerhard2002}
{Gerhard}, O., {Arnaboldi}, M., {Freeman}, K.~C., \& {Okamura}, S. 2002, \apjl,
  580, L121, \dodoi{10.1086/345657}

\bibitem[{{Giodini} {et~al.}(2009){Giodini}, {Pierini}, {Finoguenov}, {Pratt},
  {Boehringer}, {Leauthaud}, {Guzzo}, {Aussel}, {Bolzonella}, {Capak}, {Elvis},
  {Hasinger}, {Ilbert}, {Kartaltepe}, {Koekemoer}, {Lilly}, {Massey},
  {McCracken}, {Rhodes}, {Salvato}, {Sanders}, {Scoville}, {Sasaki}, {Smolcic},
  {Taniguchi}, {Thompson}, \& {COSMOS Collaboration}}]{giodini2009}
{Giodini}, S., {Pierini}, D., {Finoguenov}, A., {et~al.} 2009, \apj, 703, 982,
  \dodoi{10.1088/0004-637X/703/1/982}

\bibitem[{{Gonzalez} {et~al.}(2005){Gonzalez}, {Zabludoff}, \&
  {Zaritsky}}]{gonzalez2005}
{Gonzalez}, A.~H., {Zabludoff}, A.~I., \& {Zaritsky}, D. 2005, \apj, 618, 195,
  \dodoi{10.1086/425896}

\bibitem[{{Gonz{\'a}lez} {et~al.}(2010){Gonz{\'a}lez}, {Labb{\'e}}, {Bouwens},
  {Illingworth}, {Franx}, {Kriek}, \& {Brammer}}]{gonzalez2010}
{Gonz{\'a}lez}, V., {Labb{\'e}}, I., {Bouwens}, R.~J., {et~al.} 2010, \apj,
  713, 115, \dodoi{10.1088/0004-637X/713/1/115}

\bibitem[{{Gullieuszik} {et~al.}(2020){Gullieuszik}, {Poggianti}, {McGee},
  {Moretti}, {Vulcani}, {Tonnesen}, {Roediger}, {Jaff{\'e}}, {Fritz},
  {Franchetto}, {Omizzolo}, {Bettoni}, {Radovich}, \& {Wolter}}]{gull2020}
{Gullieuszik}, M., {Poggianti}, B.~M., {McGee}, S.~L., {et~al.} 2020, \apj,
  899, 13, \dodoi{10.3847/1538-4357/aba3cb}

\bibitem[{{Henriques} {et~al.}(2019){Henriques}, {White}, {Lilly}, {Bell},
  {Bluck}, \& {Terrazas}}]{henriques2019}
{Henriques}, B. M.~B., {White}, S. D.~M., {Lilly}, S.~J., {et~al.} 2019,
  \mnras, 485, 3446, \dodoi{10.1093/mnras/stz577}

\bibitem[{{Hlavacek-Larrondo} {et~al.}(2020){Hlavacek-Larrondo}, {Rhea},
  {Webb}, {McDonald}, {Muzzin}, {Wilson}, {Finner}, {Valin}, {Bonaventura},
  {Cooper}, {Fabian}, {Gendron-Marsolais}, {Jee}, {Lidman}, {Mezcua}, {Noble},
  {Russell}, {Surace}, {Trudeau}, \& {Yee}}]{hl2020}
{Hlavacek-Larrondo}, J., {Rhea}, C.~L., {Webb}, T., {et~al.} 2020, \apjl, 898,
  L50, \dodoi{10.3847/2041-8213/ab9ca5}

\bibitem[{{Hogg} {et~al.}(2002){Hogg}, {Baldry}, {Blanton}, \&
  {Eisenstein}}]{hogg2002}
{Hogg}, D.~W., {Baldry}, I.~K., {Blanton}, M.~R., \& {Eisenstein}, D.~J. 2002,
  arXiv e-prints, astro.
\newblock \doarXiv{astro-ph/0210394}

\bibitem[{{Kapferer} {et~al.}(2009){Kapferer}, {Sluka}, {Schindler}, {Ferrari},
  \& {Ziegler}}]{kapferer2009}
{Kapferer}, W., {Sluka}, C., {Schindler}, S., {Ferrari}, C., \& {Ziegler}, B.
  2009, \aap, 499, 87, \dodoi{10.1051/0004-6361/200811551}

\bibitem[{{Kennicutt}(1998)}]{k1998}
{Kennicutt}, Robert~C., J. 1998, \araa, 36, 189,
  \dodoi{10.1146/annurev.astro.36.1.189}

\bibitem[{{Krick} \& {Bernstein}(2007)}]{kb2007}
{Krick}, J.~E., \& {Bernstein}, R.~A. 2007, \aj, 134, 466,
  \dodoi{10.1086/518787}

\bibitem[{{Kronberger} {et~al.}(2008){Kronberger}, {Kapferer}, {Ferrari},
  {Unterguggenberger}, \& {Schindler}}]{kronberger2008}
{Kronberger}, T., {Kapferer}, W., {Ferrari}, C., {Unterguggenberger}, S., \&
  {Schindler}, S. 2008, \aap, 481, 337, \dodoi{10.1051/0004-6361:20078904}

\bibitem[{{Lee} {et~al.}(2000){Lee}, {Richer}, \& {McCall}}]{lee2000}
{Lee}, H., {Richer}, M.~G., \& {McCall}, M.~L. 2000, \apjl, 530, L17,
  \dodoi{10.1086/312483}

\bibitem[{{Lidman} {et~al.}(2012){Lidman}, {Suherli}, {Muzzin}, {Wilson},
  {Demarco}, {Brough}, {Rettura}, {Cox}, {DeGroot}, {Yee}, {Gilbank},
  {Hoekstra}, {Balogh}, {Ellingson}, {Hicks}, {Nantais}, {Noble}, {Lacy},
  {Surace}, \& {Webb}}]{lidman2012}
{Lidman}, C., {Suherli}, J., {Muzzin}, A., {et~al.} 2012, \mnras, 427, 550,
  \dodoi{10.1111/j.1365-2966.2012.21984.x}

\bibitem[{{Lin} {et~al.}(2013){Lin}, {Brodwin}, {Gonzalez}, {Bode},
  {Eisenhardt}, {Stanford}, \& {Vikhlinin}}]{lin2013}
{Lin}, Y.-T., {Brodwin}, M., {Gonzalez}, A.~H., {et~al.} 2013, \apj, 771, 61,
  \dodoi{10.1088/0004-637X/771/1/61}

\bibitem[{{Lonsdale} {et~al.}(2003){Lonsdale}, {Lonsdale}, {Smith}, \&
  {Diamond}}]{lonsdale2003}
{Lonsdale}, C.~J., {Lonsdale}, C.~J., {Smith}, H.~E., \& {Diamond}, P.~J. 2003,
  \apj, 592, 804, \dodoi{10.1086/375778}

\bibitem[{{McCarthy} {et~al.}(2010){McCarthy}, {Schaye}, {Ponman}, {Bower},
  {Booth}, {Dalla Vecchia}, {Crain}, {Springel}, {Theuns}, \&
  {Wiersma}}]{mccarthy2010}
{McCarthy}, I.~G., {Schaye}, J., {Ponman}, T.~J., {et~al.} 2010, \mnras, 406,
  822, \dodoi{10.1111/j.1365-2966.2010.16750.x}

\bibitem[{{McDonald} {et~al.}(2016){McDonald}, {Stalder}, {Bayliss}, {Allen},
  {Applegate}, {Ashby}, {Bautz}, {Benson}, {Bleem}, {Brodwin}, {Carlstrom},
  {Chiu}, {Desai}, {Gonzalez}, {Hlavacek-Larrondo}, {Holzapfel}, {Marrone},
  {Miller}, {Reichardt}, {Saliwanchik}, {Saro}, {Schrabback}, {Stanford},
  {Stark}, {Vieira}, \& {Zenteno}}]{mcdonald2016}
{McDonald}, M., {Stalder}, B., {Bayliss}, M., {et~al.} 2016, \apj, 817, 86,
  \dodoi{10.3847/0004-637X/817/2/86}

\bibitem[{{McDonald} {et~al.}(2019){McDonald}, {McNamara}, {Voit}, {Bayliss},
  {Benson}, {Brodwin}, {Canning}, {Florian}, {Garmire}, {Gaspari}, {Gladders},
  {Hlavacek-Larrondo}, {Kara}, {Reichardt}, {Russell}, {Saro}, {Sharon},
  {Somboonpanyakul}, {Tremblay}, \& {van Weeren}}]{mcdonald2019}
{McDonald}, M., {McNamara}, B.~R., {Voit}, G.~M., {et~al.} 2019, \apj, 885, 63,
  \dodoi{10.3847/1538-4357/ab464c}

\bibitem[{{McMullin} {et~al.}(2007){McMullin}, {Waters}, {Schiebel}, {Young},
  \& {Golap}}]{casa}
{McMullin}, J.~P., {Waters}, B., {Schiebel}, D., {Young}, W., \& {Golap}, K.
  2007, in Astronomical Society of the Pacific Conference Series, Vol. 376,
  Astronomical Data Analysis Software and Systems XVI, ed. R.~A. {Shaw},
  F.~{Hill}, \& D.~J. {Bell}, 127

\bibitem[{{McNamara} \& {Nulsen}(2012)}]{mcnamara2012}
{McNamara}, B.~R., \& {Nulsen}, P.~E.~J. 2012, New Journal of Physics, 14,
  055023, \dodoi{10.1088/1367-2630/14/5/055023}

\bibitem[{{Muzzin} {et~al.}(2009){Muzzin}, {Wilson}, {Yee}, {Hoekstra},
  {Gilbank}, {Surace}, {Lacy}, {Blindert}, {Majumdar}, {Demarco}, {Gardner},
  {Gladders}, \& {Lonsdale}}]{muzzin2009}
{Muzzin}, A., {Wilson}, G., {Yee}, H.~K.~C., {et~al.} 2009, \apj, 698, 1934,
  \dodoi{10.1088/0004-637X/698/2/1934}

\bibitem[{{Narayanan} \& {Krumholz}(2014)}]{narayanan2014}
{Narayanan}, D., \& {Krumholz}, M.~R. 2014, \mnras, 442, 1411,
  \dodoi{10.1093/mnras/stu834}

\bibitem[{{Oliva-Altamirano} {et~al.}(2014){Oliva-Altamirano}, {Brough},
  {Lidman}, {Couch}, {Hopkins}, {Colless}, {Taylor}, {Robotham},
  {Gunawardhana}, {Ponman}, {Baldry}, {Bauer}, {Bland-Hawthorn}, {Cluver},
  {Cameron}, {Conselice}, {Driver}, {Edge}, {Graham}, {van Kampen},
  {Lara-L{\'o}pez}, {Liske}, {L{\'o}pez-S{\'a}nchez}, {Loveday}, {Mahajan},
  {Peacock}, {Phillipps}, {Pimbblet}, \& {Sharp}}]{oa2014}
{Oliva-Altamirano}, P., {Brough}, S., {Lidman}, C., {et~al.} 2014, \mnras, 440,
  762, \dodoi{10.1093/mnras/stu277}

\bibitem[{{Pope} {et~al.}(2008){Pope}, {Chary}, {Alexander}, {Armus},
  {Dickinson}, {Elbaz}, {Frayer}, {Scott}, \& {Teplitz}}]{pope2008}
{Pope}, A., {Chary}, R.-R., {Alexander}, D.~M., {et~al.} 2008, \apj, 675, 1171,
  \dodoi{10.1086/527030}

\bibitem[{{Puchwein} {et~al.}(2008){Puchwein}, {Sijacki}, \&
  {Springel}}]{puchwein2008}
{Puchwein}, E., {Sijacki}, D., \& {Springel}, V. 2008, \apjl, 687, L53,
  \dodoi{10.1086/593352}

\bibitem[{{Puchwein} {et~al.}(2010){Puchwein}, {Springel}, {Sijacki}, \&
  {Dolag}}]{puchwein2010}
{Puchwein}, E., {Springel}, V., {Sijacki}, D., \& {Dolag}, K. 2010, \mnras,
  406, 936, \dodoi{10.1111/j.1365-2966.2010.16786.x}

\bibitem[{{Salpeter}(1955)}]{salpeter1955}
{Salpeter}, E.~E. 1955, \apj, 121, 161, \dodoi{10.1086/145971}

\bibitem[{{Sand} {et~al.}(2011){Sand}, {Graham}, {Bildfell}, {Foley},
  {Pritchet}, {Zaritsky}, {Hoekstra}, {Just}, {Herbert-Fort}, \&
  {Sivanandam}}]{sand2011}
{Sand}, D.~J., {Graham}, M.~L., {Bildfell}, C., {et~al.} 2011, \apj, 729, 142,
  \dodoi{10.1088/0004-637X/729/2/142}

\bibitem[{{Schulz} \& {Struck}(2001)}]{ss2001}
{Schulz}, S., \& {Struck}, C. 2001, \mnras, 328, 185,
  \dodoi{10.1046/j.1365-8711.2001.04847.x}

\bibitem[{{Shipley} {et~al.}(2018){Shipley}, {Lange-Vagle}, {Marchesini},
  {Brammer}, {Ferrarese}, {Stefanon}, {Kado-Fong}, {Whitaker}, {Oesch},
  {Feinstein}, {Labb{\'e}}, {Lundgren}, {Martis}, {Muzzin}, {Nedkova},
  {Skelton}, \& {van der Wel}}]{Shipley2018}
{Shipley}, H.~V., {Lange-Vagle}, D., {Marchesini}, D., {et~al.} 2018, \apjs,
  235, 14, \dodoi{10.3847/1538-4365/aaacce}

\bibitem[{{Sijacki} {et~al.}(2007){Sijacki}, {Springel}, {Di Matteo}, \&
  {Hernquist}}]{sijacki2007}
{Sijacki}, D., {Springel}, V., {Di Matteo}, T., \& {Hernquist}, L. 2007,
  \mnras, 380, 877, \dodoi{10.1111/j.1365-2966.2007.12153.x}

\bibitem[{{Sivanandam} {et~al.}(2010){Sivanandam}, {Rieke}, \&
  {Rieke}}]{sivan2010}
{Sivanandam}, S., {Rieke}, M.~J., \& {Rieke}, G.~H. 2010, \apj, 717, 147,
  \dodoi{10.1088/0004-637X/717/1/147}

\bibitem[{{Smith} {et~al.}(2010){Smith}, {Lucey}, {Hammer}, {Hornschemeier},
  {Carter}, {Hudson}, {Marzke}, {Mouhcine}, {Eftekharzadeh}, {James},
  {Khosroshahi}, {Kourkchi}, \& {Karick}}]{smith2010}
{Smith}, R.~J., {Lucey}, J.~R., {Hammer}, D., {et~al.} 2010, \mnras, 408, 1417,
  \dodoi{10.1111/j.1365-2966.2010.17253.x}

\bibitem[{{Somboonpanyakul} {et~al.}(2021){Somboonpanyakul}, {McDonald},
  {Bayliss}, {Voit}, {Donahue}, {Gaspari}, {Dahle}, {Rivera-Thorsen}, \&
  {Stark}}]{Somboonpanyakul2021}
{Somboonpanyakul}, T., {McDonald}, M., {Bayliss}, M., {et~al.} 2021, \apjl,
  907, L12, \dodoi{10.3847/2041-8213/abd540}

\bibitem[{{Sun} {et~al.}(2010){Sun}, {Donahue}, {Roediger}, {Nulsen}, {Voit},
  {Sarazin}, {Forman}, \& {Jones}}]{sun2010}
{Sun}, M., {Donahue}, M., {Roediger}, E., {et~al.} 2010, \apj, 708, 946,
  \dodoi{10.1088/0004-637X/708/2/946}

\bibitem[{{Sun} {et~al.}(2007){Sun}, {Donahue}, \& {Voit}}]{sun2007}
{Sun}, M., {Donahue}, M., \& {Voit}, G.~M. 2007, \apj, 671, 190,
  \dodoi{10.1086/522690}

\bibitem[{Tonini {et~al.}(2012)Tonini, Bernyk, Croton, Maraston, \&
  Thomas}]{Tonini_2012}
Tonini, C., Bernyk, M., Croton, D., Maraston, C., \& Thomas, D. 2012, The
  Astrophysical Journal, 759, 43, \dodoi{10.1088/0004-637x/759/1/43}

\bibitem[{{Trudeau} {et~al.}(2019){Trudeau}, {Webb}, {Hlavacek-Larrondo},
  {Noble}, {Gendron-Marsolais}, {Lidman}, {Mezcua}, {Muzzin}, {Wilson}, \&
  {Yee}}]{trudeau2019}
{Trudeau}, A., {Webb}, T., {Hlavacek-Larrondo}, J., {et~al.} 2019, \mnras, 487,
  1210, \dodoi{10.1093/mnras/stz1364}

\bibitem[{{Vollmer} {et~al.}(2001){Vollmer}, {Cayatte}, {Balkowski}, \&
  {Duschl}}]{vollmer2001}
{Vollmer}, B., {Cayatte}, V., {Balkowski}, C., \& {Duschl}, W.~J. 2001, \apj,
  561, 708, \dodoi{10.1086/323368}

\bibitem[{Webb {et~al.}(2015)Webb, Noble, DeGroot, Wilson, Muzzin, Bonaventura,
  Cooper, Delahaye, Foltz, Lidman, \& et~al.}]{Webb_2015}
Webb, T., Noble, A., DeGroot, A., {et~al.} 2015, The Astrophysical Journal,
  809, 173, \dodoi{10.1088/0004-637x/809/2/173}

\bibitem[{{Webb} {et~al.}(2015){Webb}, {Muzzin}, {Noble}, {Bonaventura},
  {Geach}, {Hezaveh}, {Lidman}, {Wilson}, {Yee}, {Surace}, \&
  {Shupe}}]{webb2015a}
{Webb}, T. M.~A., {Muzzin}, A., {Noble}, A., {et~al.} 2015, \apj, 814, 96,
  \dodoi{10.1088/0004-637X/814/2/96}

\bibitem[{{Webb} {et~al.}(2017){Webb}, {Lowenthal}, {Yun}, {Noble}, {Muzzin},
  {Wilson}, {Yee}, {Cybulski}, {Aretxaga}, \& {Hughes}}]{webb2017}
{Webb}, T. M.~A., {Lowenthal}, J., {Yun}, M., {et~al.} 2017, \apjl, 844, L17,
  \dodoi{10.3847/2041-8213/aa7749}

\bibitem[{{Williams} {et~al.}(2007){Williams}, {Ciardullo}, {Durrell},
  {Vinciguerra}, {Feldmeier}, {Jacoby}, {Sigurdsson}, {von Hippel}, {Ferguson},
  {Tanvir}, {Arnaboldi}, {Gerhard}, {Aguerri}, \& {Freeman}}]{williams2007}
{Williams}, B.~F., {Ciardullo}, R., {Durrell}, P.~R., {et~al.} 2007, \apj, 656,
  756, \dodoi{10.1086/510149}

\bibitem[{{Wilson} {et~al.}(2009){Wilson}, {Muzzin}, {Yee}, {Lacy}, {Surace},
  {Gilbank}, {Blindert}, {Hoekstra}, {Majumdar}, {Demarco}, {Gardner},
  {Gladders}, \& {Lonsdale}}]{wilson2009}
{Wilson}, G., {Muzzin}, A., {Yee}, H.~K.~C., {et~al.} 2009, \apj, 698, 1943,
  \dodoi{10.1088/0004-637X/698/2/1943}

\bibitem[{{Yoshida} {et~al.}(2008){Yoshida}, {Yagi}, {Komiyama}, {Furusawa},
  {Kashikawa}, {Koyama}, {Yamanoi}, {Hattori}, \& {Okamura}}]{yoshida2008}
{Yoshida}, M., {Yagi}, M., {Komiyama}, Y., {et~al.} 2008, \apj, 688, 918,
  \dodoi{10.1086/592430}

\bibitem[{{Zibetti} {et~al.}(2009){Zibetti}, {Charlot}, \& {Rix}}]{zibetti2009}
{Zibetti}, S., {Charlot}, S., \& {Rix}, H.-W. 2009, \mnras, 400, 1181,
  \dodoi{10.1111/j.1365-2966.2009.15528.x}

\bibitem[{{Zibetti} {et~al.}(2005){Zibetti}, {White}, {Schneider}, \&
  {Brinkmann}}]{zibetti2005}
{Zibetti}, S., {White}, S. D.~M., {Schneider}, D.~P., \& {Brinkmann}, J. 2005,
  \mnras, 358, 949, \dodoi{10.1111/j.1365-2966.2005.08817.x}

\end{thebibliography}
\bibliographystyle{aasjournal}



\end{document}